\documentclass[a4paper,11pt]{article}
\usepackage{pos}
\usepackage[textsize=tiny]{todonotes}
\usepackage{wrapfig}
\usepackage{natbib}
\usepackage{lineno}
\usepackage{setspace}
\usepackage{textcomp}
\title{Recent Progress in Solar Atmospheric Neutrino Searches with IceCube}
\ShortTitle{Solar Atmospheric Neutrinos with IceCube} 

\author{The IceCube Collaboration \\{\normalsize \normalfont(a complete list of authors can be found at the end of the proceedings)}}

\emailAdd{jvillarreal@icecube.wisc.edu}
\emailAdd{gerrit.roellinghoff@icecube.wisc.edu}

\abstract{
    Cosmic-rays interacting with nucleons in the solar atmosphere produce a cascade of particles that give rise to a flux of high-energy neutrinos and gamma-rays. Fermi has observed this gamma-ray flux; however, the associated neutrino flux has escaped observation. In this contribution, we put forward two strategies to detect these neutrinos, which, if seen, would push forward our understanding of the solar atmosphere and provide a new testing ground of neutrino properties. First, we will extend the previous analysis, which used high-energy through-going muon events collected in the years of maximum solar activity and yielded only flux upper limits, to include data taken during the solar minimum from 2018 to 2020. Extending the analysis to the solar minimum is important as the gamma-ray data collected during past solar cycles indicates a possible enhancement in the high-energy neutrino flux. Second, we will incorporate sub-TeV events and include contributions from all neutrino flavors. These will improve our analysis sensitivity since the solar atmospheric spectrum is soft and, due to oscillation, contains significant contributions of all neutrino flavors. As we will present in this contribution, these complementary strategies yield a significant improvement in sensitivity, making substantial progress towards observing this flux.

\vspace{4mm}
{\bfseries Corresponding authors:}
Josh Villarreal$^{1*}$, Gerrit Roellinghoff$^{2}$, Jeffrey Lazar$^{1,3}$\\
{$^{1}$ \itshape Department of Physics and Laboratory for Particle Physics and Cosmology, Harvard University, Cambridge, MA 02138, USA}\\
{$^{2}$ \itshape Dept. of Physics, Sungkyunkwan University, Suwon 16419, Korea}\\
{$^{3}$ \itshape Dept. of Physics and Wisconsin IceCube Particle Astrophysics Center, University of Wisconsin–Madison, Madison, WI 53706, USA}\\[4mm]
$^*$ Presenter
}
\FullConference{37$^{\rm{th}}$ International Cosmic Ray Conference (ICRC 2021)\\
		July 12th -- 23rd, 2021\\
		Online -- Berlin, Germany}

\begin{document}
\maketitle

\section{Introduction}
Humans have observed and studied the Sun, our closest star, since prehistoric times.
Recently, the Sun has been observed in neutrinos, an accomplishment achieved first by the SuperKamiokande collaboration and subsequently by the SNO and Borexino collaborations \cite{kamiokandecollaboration2016solar, martin2009results, borexino2020}. These collaborations observed the MeV-scale neutrinos produced in the nuclear processes which power the Sun.

At higher energies, $\gtrsim$10~GeV, there is another well-predicted but yet unmeasured flux of solar neutrinos.
These neutrinos are the partners of the gamma rays from the solar disk, and are produced from decaying hadrons, which are produced when cosmic-rays interact with solar matter.
Since the mechanism by which these neutrinos are created is the same as those produced in Earth's atmosphere, they are commonly called \textit{solar atmospheric neutrinos}.
While the fluxes of solar and conventional atmospheric neutrinos are at comparable levels, the rate of solar atmospheric neutrinos at the detector is suppressed relative to their conventional counterparts by the fractional solid angle of the Sun, $\Omega_{\odot}/4\pi\simeq 5.4\times 10^{-6}$.
Because of this suppression, solar atmospheric neutrinos have so far remained undetected.

The detection of this flux remains an important scientific goal for a number of reasons.
Current data shows an enhancement of the solar disk gamma-ray flux relative to the nominal flux expectation \cite{Abdo2011}.
One explanation that has been put forward to resolve this phenomenon is that the solar magnetic field may redirect primary cosmic rays towards Earth, thus increasing the observed solar gamma-ray flux.
Since solar atmospheric neutrinos are produced in the same hadronic interactions, measurement of the neutrino flux will shed light on this problem.
Additionally, these neutrinos are an irreducible background to dark matter indirect detection searches which look for neutrinos created when dark matter annihilates in the solar core.
As such, they result in a floor on the dark matter nucleon cross section, below which experimental probing becomes challenging.
Furthermore, with sufficient statistics, this flux may give us a new baseline with which to perform oscillation analyses, since the first oscillation maximum comes at 10 TeV for the Sun-to-Earth distance.
While current generation experiments likely cannot accumulate enough statistics to realize such an analysis, knowledge of the flux level will inform future experiments.
Thus, first measurements which set the normalization of this spectrum are important for numerous scientific goals.

The IceCube Neutrino Telescope is a gigaton-scale detector located in the ice between 1450~m and 2450~m beneath the geographic South Pole.
The ice is instrumented with 5160 digital optical modules (DOMs) housing a photomultiplier tube and on-board digitization hardware.
DOMs can detect Cherenkov radiation emitted by charged particles (created by neutrino interactions) traveling through the ice, offering a way to reconstruct direction, energy and flavor of neutrinos interacting with the ice or surrounding bedrock \cite{detectorjinst}.


In this proceeding, we present the current status of IceCube's efforts to detect the solar atmospheric neutrino flux.
In Sec.~\ref{sec:sig_and_bg}, we elaborate on theoretical flux predictions for the solar atmospheric neutrino signal and its associated backgrounds.
In Sec.~\ref{sec:analysis_methods}, we describe the methodology of our current approach of building and improving on a preceding analysis with IceCube, showing associated preliminary sensitivities to reference fluxes from theoretical predictions in Sec.~\ref{sec:results}.
In Sec.~\ref{sec:new_selection} we detail the development of a new low- and medium energy event selection designed to improve sensitivities by incorporating all-flavor events and cascades in addition to the muon neutrino tracks in the high-energy selection.

\section{Signal and Background}
\label{sec:sig_and_bg}
\begin{figure}[t]
    \centering
    \includegraphics[width=\textwidth]{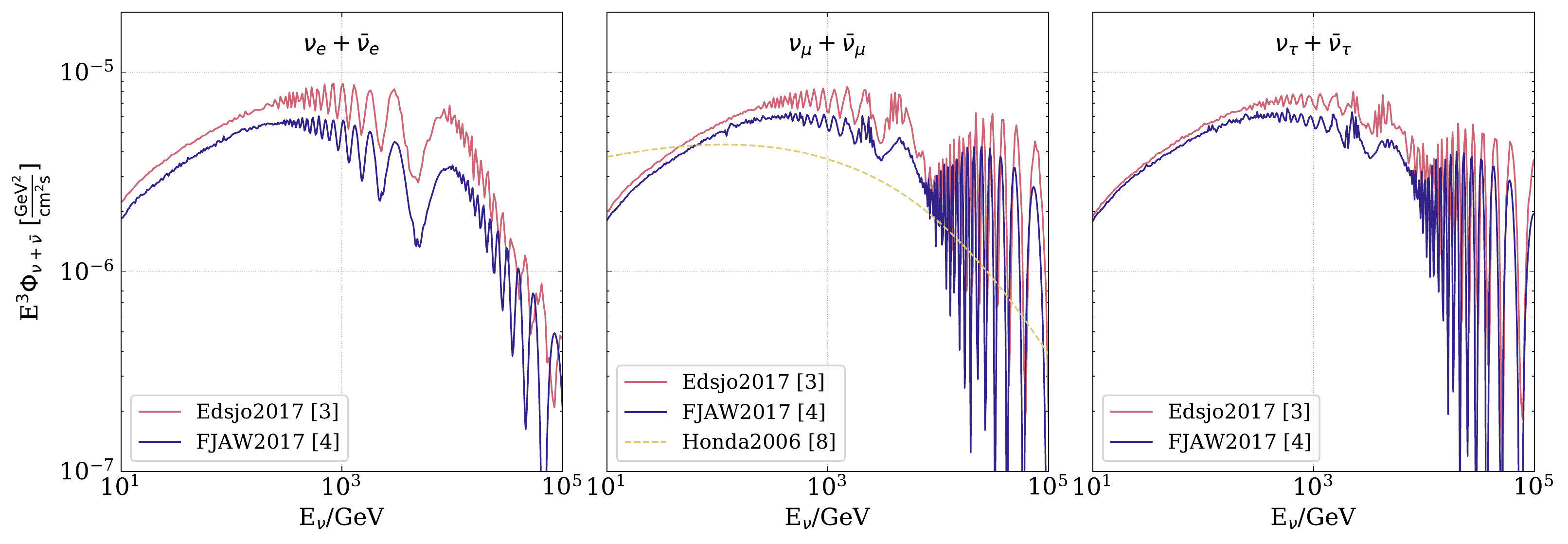}
    \caption{Neutrino flux after propagating 1 AU from the sun for our reference fluxes from \cite{Edsjo2017} (Edsjo2017) and \cite{FJAW2017} (FJAW2017). For Edsjo2017, we pick the calculation based on Hillas-Gaisser H3a \cite{Gaisser2012} as the Cosmic Ray model, Serenelli+Stein \cite{Serenelli2009,Stein1998} as the solar density model and normal neutrino mass ordering. For FJAW2017, we pick the calculation based on Hillas-Gaisser H4a \cite{Gaisser2012} as the cosmic ray model and FJAW2017's custom sun model \cite{FJAW2017}. For $\mu$-neutrinos, we also compare the fluxes to the expected flux for conventional atmospheric neutrinos from the direction of the solar disk using a flux prediction from \cite{Honda2007}.}
    \label{pic:flux_comp}
\end{figure}
When high-energy cosmic rays interact with the solar atmosphere, the resulting hadronic cascades produce secondary particles that can decay into gamma-rays and neutrinos, similar to the creation of the conventional atmospheric neutrino flux on Earth. A first prediction of the resulting particle fluxes was made in 1991 \cite{Seckel1991}, with further methodological improvement by using Monte Carlo methods made in 1996 \cite{IT1996}, resulting in a theoretical flux roughly following $E^{-3}$ until $10$ TeV before steeply falling off. Newer theoretical predictions made in recent years predict slightly lower fluxes \cite{FJAW2017,Edsjo2017,Ng2017}, but are generally in agreement within a factor of 2 with each other. Figure \ref{pic:flux_comp} shows the $\nu + \bar{\nu}$ flux predictions for a reference flux each from \cite{FJAW2017,Edsjo2017}. 

Gamma-ray data from Fermi-LAT and EGRET during different levels of solar activity \cite{Linden_2018,Orlando_2008,Abdo2011,Ng2016} suggests an anti-correlation between solar activity (determined by the number of sunspots) and high-energy gamma-ray emission, with the majority of very high-energy events occurring during the solar minimum of 2008-2010, implying a corresponding enhancement of the high-energy neutrino flux. Additionally, a possible 50\% increase in flux during times of low solar activity has been predicted \cite{Masip2018}. 

The relevant background for this signal are atmospheric muons, conventional atmospheric neutrinos and the diffuse astrophysical flux.
Atmospheric muons can be effectively vetoed by limiting the event selection to up-going events, which we do for our high energy selection, while down-going events need additional filtering to distinguish between atmospheric muons and neutrinos; see Sec.~\ref{sec:new_selection}.


\section{Analysis Method}
\label{sec:analysis_methods}
For high energy events, we use a new IceCube dataset using 9 years of 86-string data from 2011 to 2020, notably including the solar minimum of 2018-2020. This IceCube dataset features an improved energy reconstruction and angular reconstruction compared to previous samples, and all 9 years feature the full detector configuration \cite{Bellenghi:2021}. The event selection is limited to upgoing events (sun declination $\delta_{sun} > -5^{\circ}$), using the Earth as an effective atmospheric muon veto to create a high-purity neutrino sample. It uses only $\nu_{\mu} + \bar{\nu}_{\mu}$ track events, owing to their superior angular reconstruction at high-energies. 

Extending on the previous analysis, we apply an unbinned likelihood method to this event selection with our likelihood function being
\begin{equation}
    \mathcal{L}(n_s) = \prod_{i=1}^{N} \Big{[} \frac{n_S}{N}p_{sig}(\vec{\theta}_i;\phi_{sig})+(1-\dfrac{n_S}{N})p_{bkg}(\vec{\theta}_i;\phi_{atm} + \phi_{astro})\Big{]},
\end{equation}
where $\vec{\theta}$ depends on the format of the chosen probability density functions (PDFs), $p_{sig}$ and $p_{bkg}$ denote signal and background PDF respectively and $n_s$ is the amount of signal events. $N$ is the total amount of events in the event selection, $\phi_{sig}$ is the signal flux and $\phi_{atm} + \phi_{astro}$ the total background due to conventional atmospheric flux and diffuse astrophysical flux. We construct our test statistic $\lambda$ as a log-likelihood ratio, with $\hat{n}_S$ being the best-fit value for $n_S$. We allow for negative values of $n_S$ to allow for differentiation between positive and negative excess, allowing for the observation of potential shadowing effects:

\begin{equation}
    \lambda = 2 \cdot \textrm{sign}(\hat{n}_S) \cdot \ln \bigl(\dfrac{\mathcal{L}(\hat{n}_S)}{\mathcal{L}(0)} \bigr).
\end{equation}

\begin{figure}[t]
    \centering
    \includegraphics[width=0.85\textwidth]{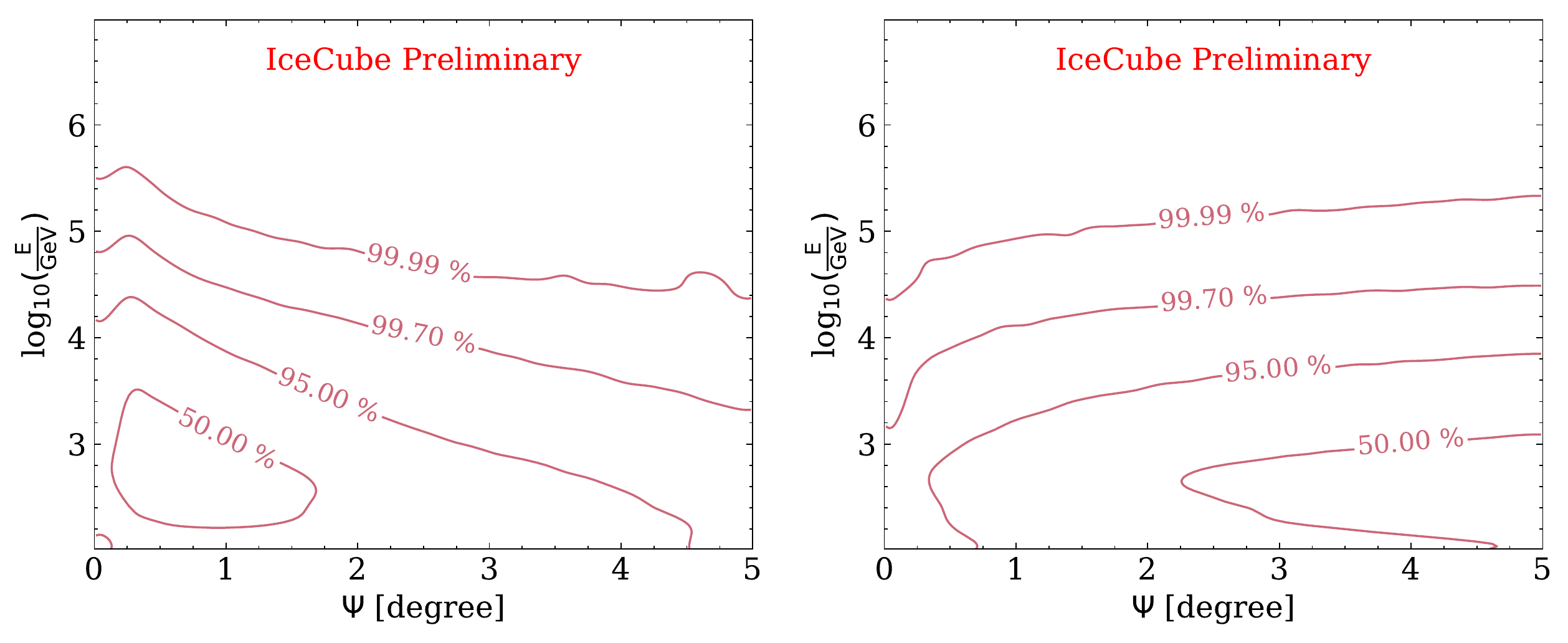}
    \caption{Probability density function contours of signal (left) and background (right) in angular distance from the sun $\Psi$ and reconstructed energy E for high-energy tracks.}
    \label{pic:cont}
\end{figure}

We have chosen our probability density functions (PDFs) for signal and background to be constructed akin to \cite{Jin2021}, with $\vec{\theta} = [\Psi, \mathrm{E}]$ (for $\Psi$ the angular distance from the Sun), by sampling Monte-Carlo events from the direction of the Sun and weighting with the corresponding flux, see Figure \ref{pic:cont}. Since we limit ourselves to up-going muon tracks, our background events stem only from the conventional atmospheric and diffuse astrophysical flux, since the Earth works as an effective atmospheric muon veto. The range of our PDFs is limited to $[0^{\circ}, 5^{\circ}] \times [10^2 \textrm{ GeV}, 10^7 \textrm{ GeV}]$, covering about $\sim$96\% of signal events.

\section{Results}
\label{sec:results}
\subsection{Previous analysis}

A previous search for solar atmospheric neutrinos was done with 7 years of IceCube data (2010-2016) taken during the austral winter, when the declination of the Sun is above $-5^{\circ}$. Using a baseline flux from \cite{Edsjo2017}, the analysis set a 90\% upper limit of $1.02^{+0.20}_{-0.18}\cdot10^{-13}$~$\mathrm{GeV^{-1}cm^{-2}s^{-1}}$ at 1~\textrm{TeV}. The analysis time period covered only times of increased solar activity. The  upper limit is an order of magnitude larger than the chosen baseline flux. \cite{Jin2021}

\subsection{Sensitivity for the new analysis}
We evaluate our sensitivities at the 90\% confidence level using the Neyman method \cite{Neyman1937}. Table \ref{table:sens} shows our integrated flux sensitivities for our reference fluxes from \cite{FJAW2017,Edsjo2017}. Depending on the flux model, the resulting sensitivities are a factor 2-4 larger than model predictions \cite{FJAW2017,Edsjo2017}.

\begin{table}[h]
\centering
\resizebox{0.8\textwidth}{!}{%
\begin{tabular}{|l|c|c|}
\hline
                           & Sens. in units of model flux & Sens. at 1 TeV [GeV$^{-1}$ cm$^{-2}$ s$^{-1}$] \\ \hline
Edsjo2017 \cite{Edsjo2017} & 2.61                         & $1.72 \cdot 10^{-14}$                          \\
FJAW2017  \cite{FJAW2017}  & 3.51                         & $1.95 \cdot 10^{-14}$                          \\ \hline
\end{tabular}%
}
\caption{Sensitivities to reference solar atmospheric flux models from \cite{FJAW2017,Edsjo2017}. Sensitivities are given in units of model flux and for the corresponding value at 1 TeV.}
\label{table:sens}
\end{table}

\begin{figure}[t]
    \centering
    \includegraphics[width=0.8\textwidth]{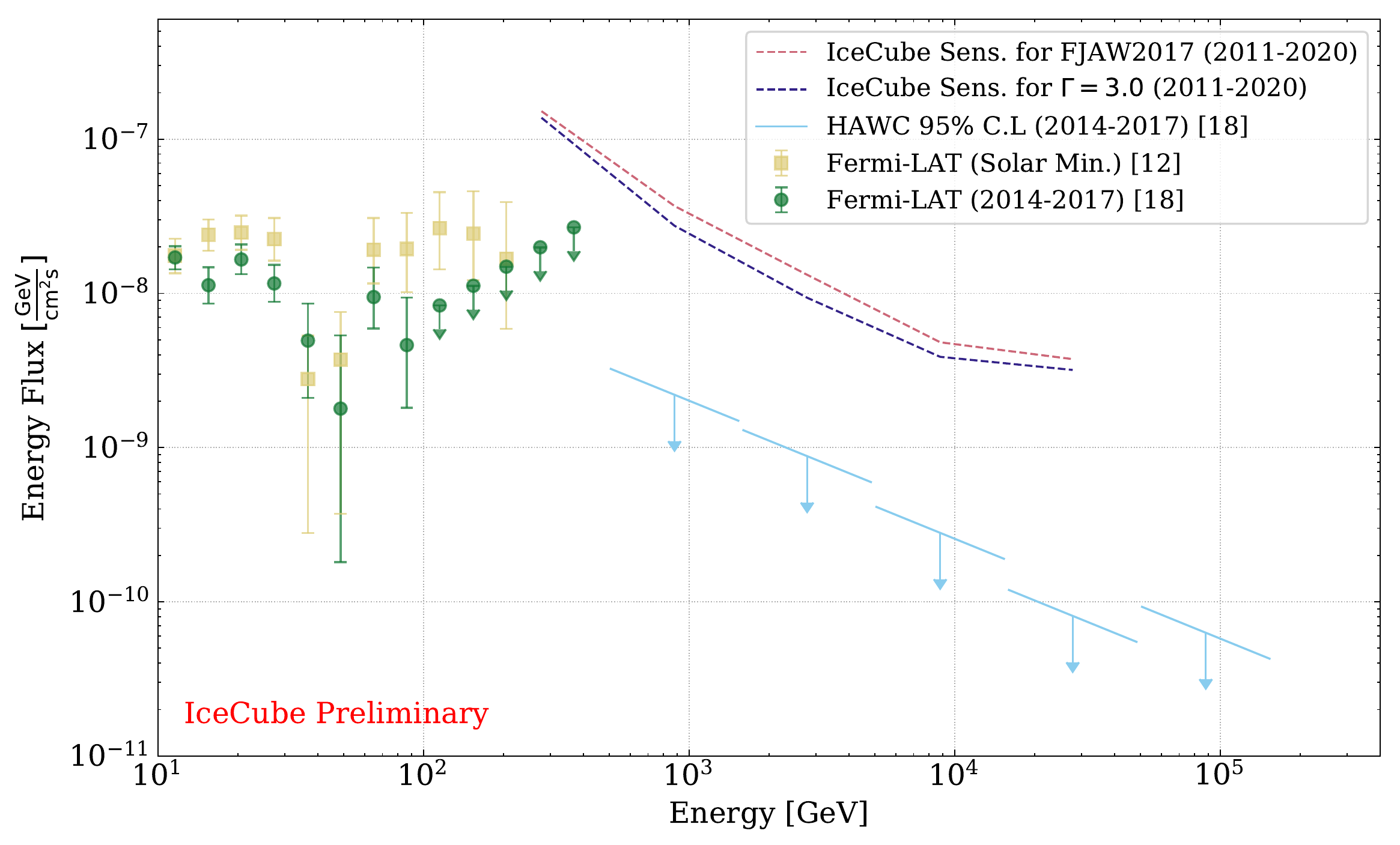}
    \caption{Differential sensitivity in half-decade energy bins for our reference flux from FJAW2017 \cite{FJAW2017} and for a power law with $\Gamma =3.0$ for 9 years of data. Also shown are gamma-ray results from HAWC \cite{Albert2018} and Fermi-LAT \cite{Linden_2018}.}
    \label{pic:diff}
\end{figure}

We also calculate the differential sensitivity in half-decade energy bins, assuming our reference flux from \cite{FJAW2017} for 9 years (2011-2020). We omit calculating the differential flux for \cite{Edsjo2017}, since shape similarity between both theoretical models means the sensitivities are nearly identical. For comparisons sake, we also show the same differential sensitivity for an assumed flux following a powerlaw with $\Gamma =3.0$. The differential sensitivities are shown in Figure \ref{pic:diff} alongside gamma-ray flux observed by Fermi-LAT \cite{Linden_2018} and upper limits provided by HAWC \cite{Albert2018}.

\section{Towards an improved event selection}
\label{sec:new_selection}
To improve the sensitivity of our solar atmospheric neutrino analysis, we are currently extending the event selection to the low- and medium-energy regimes and including all neutrino flavors and event morphologies (cascades and tracks)to do this, we are using a previously-built low-energy selection (<300~GeV) and a medium energy selection (100-600~GeV), which will  bridge the gap between low- and high-energy subselections. After removing duplicate events, we aim to combine all 3 event selections in a likelihood fit:
\begin{equation}
    \mathcal{L}_{\rm{tot}} = \mathcal{L}_{\rm{LE}}\,\mathcal{L}_{\rm{ME}}\,\mathcal{L}_{\rm{HE}},
\end{equation}
enabled by the fact that all event selections are statistically independent after removing duplicates.
\subsection{Low-Energy Selection}
\label{subsec:lo_selection}

Located within the IceCube instrumented volume is a more densely packed sub-detector known as DeepCore.
In DeepCore, the inter-string distance ranges from 42~m to 72~m, while the vertical DOM spacing ranges from 7~m to 10~m.
This higher density of DOMs allows DeepCore to detect neutrinos with energies as low as a few GeV.

OscNext is a suite of atmospheric neutrino oscillation analyses which use DeepCore data from 2011-2019.
These analyses are joined by a common event selection whose purpose is to achieve a neutrino-dominated event sample by rejecting a prevailing background from atmospheric muons and detector noise.

This event selection contains neutrinos with reconstructed energies ranging from 5~GeV to 300~GeV, and contains all neutrino flavors.
At the final level, the sample has a nominal neutrino rate of 0.991~mHz, a nominal muon rate of 0.034~mHz, and a nominal noise rate below 1~{\textmu}Hz.
The selection obtains this level of purity using a series of selection cuts combining traditional straight cuts and boosted decision trees (BDTs).
Additionally, this selection uses a BDT to differentiate between cascades and tracks.
We are also investigating the effect of relaxing this selection's cuts on our sensitivity.
Since this analysis amounts to looking for a point source, we may be able to tolerate more background than the oscNext analyses.

Adding the low-energy selection in its current form to our analysis leads to a sensitivity increase of around 20\% ($3.51 \rightarrow 2.91$ in units of model flux for \cite{FJAW2017}). 

\subsection{Medium-Energy selection}
\label{subsec:me_selection}

\begin{figure}[t]
    \centering
    \begin{minipage}{0.45\textwidth}
        \centering
        \includegraphics[width=\textwidth]{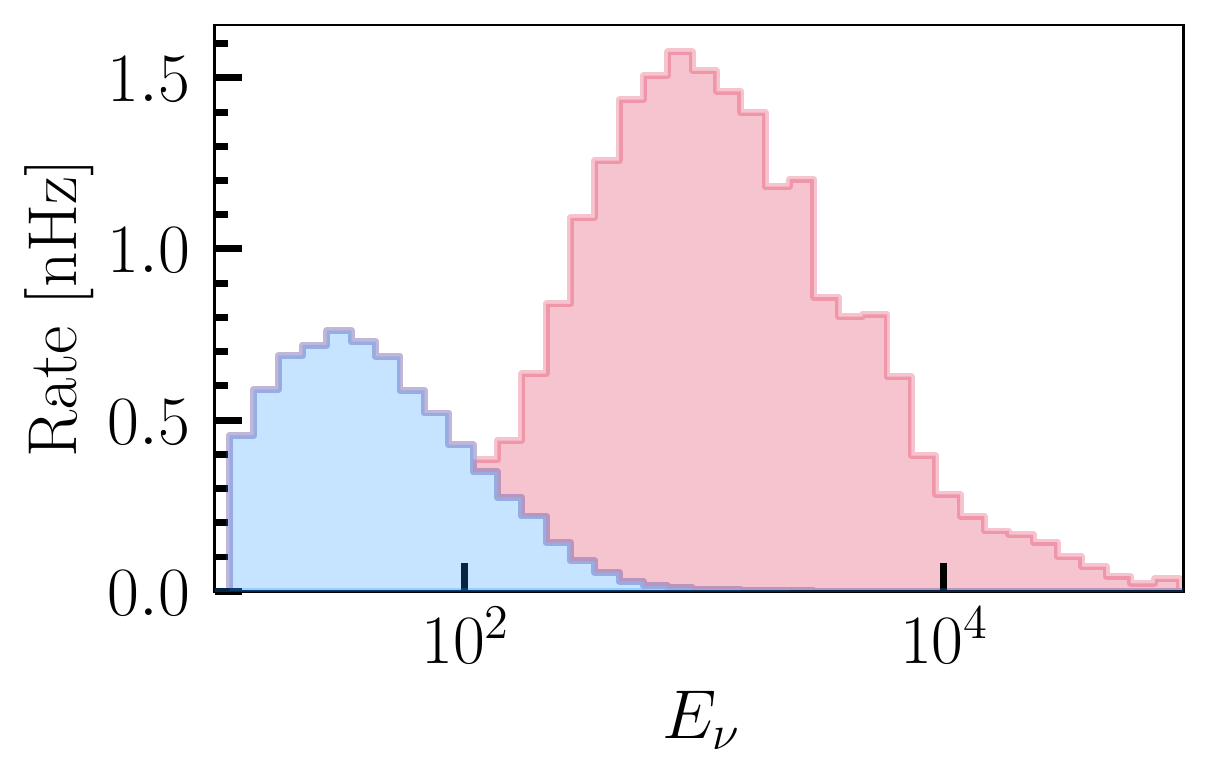}
        \caption{Stacked distribution solar atmospheric neutrino rates as a function of true neutrino energy in the low-energy (blue) and high-energy (red) selections for the FJAW model. Note the gap in coverage for neutrinos with energies in the range of 80~GeV to 300~GeV.}
        \label{fig:stacked_rate}
    \end{minipage}\hfill
    \begin{minipage}{0.45\textwidth}
        \centering
        \includegraphics[width=\textwidth]{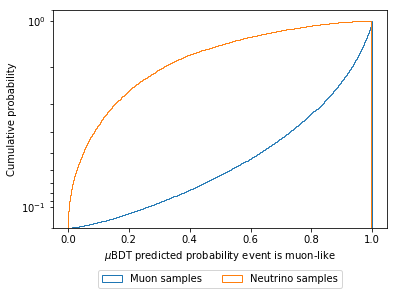}
        \caption{Cumulative muon-discriminating BDT score distributions on test sample muon and neutrino simulated datasets. Histograms are normalized so that each differential BDT score distribution integrates to $1$ before computing the cumulative distributions shown.}
        \label{fig:muBDT}
    \end{minipage}
\end{figure}

At around E$_{\nu}=$100~GeV, there is a gap in coverage between the low- and high-energy selections, see Fig.~\ref{fig:stacked_rate}.
Since the flux of solar atmospheric neutrinos is higher at lower energies, it is import to have coverage in this energy regime.
To do this, we employ IceCube's `LowUp Filter.'
This low-level trigger was designed to target low-energy, up-going neutrinos, originally used in IceCube's searches for neutrinos coming from dark matter annihilation in the Sun \cite{annihilate}.
Since these analyses share a source origin and target energy regime, it is natural to adapt the methods of one to the other.

After having selected events which pass the LowUp filter and having filtered events which may be in other portions of the selection, we plan to emulate the approach of the low-energy selection.
We perform computationally inexpensive reconstructions and make a conservative cut on zenith angle to filter out much of the atmospheric muon background which dominates in the southern sky.
After this cut, the data rate is sufficiently low to use a BDT to differentiate muons from neutrinos; see Fig.~\ref{fig:muBDT} for the current performance of this BDT. The BDT was trained on 75\% of available Monte Carlo muon and neutrino event data, and we performed a weighted two-sample Kolmogorov-Smirnov test to verify that both training and testing sets followed the same BDT score distribution, meaning that the BDT did not overtrain.
At this point, the data rate has been cut to a sufficiently low level to allow more computationally expensive reconstructions to be run.
We are in the process of studying different reconstructions in order to understand which will optimize our sensitivity.

Finally, we intend to make a second BDT to differentiate solar atmospheric neutrinos from conventional atmospheric neutrinos.
This is feasible due to the differences in zenith distribution and energy spectra of the two populations.
Additionally, while the flavor composition of each is the same at production, solar atmospheric neutrinos are able to oscillate into other flavors, leading to more cascade-like events in the solar atmospheric population.
To exploit this latter fact, we will include metrics which are tied to particle identification in this BDT, in addition to directional and energy estimates.
Since the result of this BDT is unlikely to provide clear distinction between the two populations, we plan to let the output of it enter as an analysis variable.

\section{Conclusion and Outlooks}
\label{sec:conclusion_and_outlooks}
We showed preliminary sensitivities for the solar atmospheric neutrino flux with 9 years of up-going high-energy muon neutrino event selection from IceCube. We see that they are close to the expected flux values for IceCube up to a factor of 2-4, depending on the assumed flux model. We also introduced an additional event selection aimed at improving sensitivities by including all-flavor low-energy events and described the approach for a further event selection bridging the gap between the low- and high-energy regimes. Preliminary results indicate a 20\% improvement in sensitivity by adding our low-energy all-flavor selection. These additional event selections and future improvements in analysis methods, e.g. incorporating the solar shadow and adding more variables to the likelihood fit, could put current solar atmospheric flux models within the reach of IceCube.

\bibliographystyle{JHEP}
\bibliography{solatmnu}

\clearpage
\section*{Full Author List: IceCube Collaboration}




\scriptsize
\noindent
R. Abbasi$^{17}$,
M. Ackermann$^{59}$,
J. Adams$^{18}$,
J. A. Aguilar$^{12}$,
M. Ahlers$^{22}$,
M. Ahrens$^{50}$,
C. Alispach$^{28}$,
A. A. Alves Jr.$^{31}$,
N. M. Amin$^{42}$,
R. An$^{14}$
% ...
\\

\noindent
% ...
$^{12}$ Universit{\'e} Libre de Bruxelles, Science Faculty CP230, B-1050 Brussels, Belgium \\
% ...
$^{14}$ Department of Physics and Laboratory for Particle Physics and Cosmology, Harvard University, Cambridge, MA 02138, USA \\
% ...
$^{17}$ Department of Physics, Loyola University Chicago, Chicago, IL 60660, USA \\
$^{18}$ Dept. of Physics and Astronomy, University of Canterbury, Private Bag 4800, Christchurch, New Zealand \\
% ...
$^{22}$ Niels Bohr Institute, University of Copenhagen, DK-2100 Copenhagen, Denmark \\
% ...
$^{28}$ D{\'e}partement de physique nucl{\'e}aire et corpusculaire, Universit{\'e} de Gen{\`e}ve, CH-1211 Gen{\`e}ve, Switzerland \\
% ...
$^{31}$ Karlsruhe Institute of Technology, Institute for Astroparticle Physics, D-76021 Karlsruhe, Germany  \\
% ...
$^{42}$ Bartol Research Institute and Dept. of Physics and Astronomy, University of Delaware, Newark, DE 19716, USA \\
% ...
$^{50}$ Oskar Klein Centre and Dept. of Physics, Stockholm University, SE-10691 Stockholm, Sweden \\
% ...
$^{59}$ DESY, D-15738 Zeuthen, Germany \\
% ...

\subsection*{Acknowledgements}

\noindent
USA {\textendash} U.S. National Science Foundation-Office of Polar Programs,
U.S. National Science Foundation-Physics Division,
U.S. National Science Foundation-EPSCoR,
Wisconsin Alumni Research Foundation,
Center for High Throughput Computing (CHTC) at the University of Wisconsin{\textendash}Madison,
Open Science Grid (OSG),
Extreme Science and Engineering Discovery Environment (XSEDE),
Frontera computing project at the Texas Advanced Computing Center,
U.S. Department of Energy-National Energy Research Scientific Computing Center,
Particle astrophysics research computing center at the University of Maryland,
Institute for Cyber-Enabled Research at Michigan State University,
and Astroparticle physics computational facility at Marquette University;
Belgium {\textendash} Funds for Scientific Research (FRS-FNRS and FWO),
FWO Odysseus and Big Science programmes,
and Belgian Federal Science Policy Office (Belspo);
Germany {\textendash} Bundesministerium f{\"u}r Bildung und Forschung (BMBF),
Deutsche Forschungsgemeinschaft (DFG),
Helmholtz Alliance for Astroparticle Physics (HAP),
Initiative and Networking Fund of the Helmholtz Association,
Deutsches Elektronen Synchrotron (DESY),
and High Performance Computing cluster of the RWTH Aachen;
Sweden {\textendash} Swedish Research Council,
Swedish Polar Research Secretariat,
Swedish National Infrastructure for Computing (SNIC),
and Knut and Alice Wallenberg Foundation;
Australia {\textendash} Australian Research Council;
Canada {\textendash} Natural Sciences and Engineering Research Council of Canada,
Calcul Qu{\'e}bec, Compute Ontario, Canada Foundation for Innovation, WestGrid, and Compute Canada;
Denmark {\textendash} Villum Fonden and Carlsberg Foundation;
New Zealand {\textendash} Marsden Fund;
Japan {\textendash} Japan Society for Promotion of Science (JSPS)
and Institute for Global Prominent Research (IGPR) of Chiba University;
Korea {\textendash} National Research Foundation of Korea (NRF);
Switzerland {\textendash} Swiss National Science Foundation (SNSF);
United Kingdom {\textendash} Department of Physics, University of Oxford.

\end{document}